\documentclass[aps.pre,showpacs]{revtex4}
\usepackage{graphicx}% Include figure files
\usepackage{color}\def\be{\begin{equation}}

\def\ee{\end{equation}}     
\def\bfi{\begin{figure}}
\def\efi{\end{figure}}
\def\bea{\begin{eqnarray}}
\def\eea{\end{eqnarray}}

\begin{document}

\title{Large deviations, condensation, and giant response in a statistical system}

\author{Federico Corberi}

\affiliation{Dipartimento di Fisica ``E.~R. Caianiello'', and INFN, 
Gruppo Collegato di Salerno, and CNISM, Unit\`a di Salerno,Universit\`a  di Salerno, 
via Giovanni Paolo II 132, 84084 Fisciano (SA), Italy.}
\pacs{05.40.-a, 64.60.Bd}

\begin{abstract}
We study the probability distribution $P$ of the sum of a large number of
non-identically distributed random variables $n_m$. Condensation
of fluctuations, the phenomenon whereby one of such variables provides 
a macroscopic contribution to the global probability, 
is discussed and interpreted in analogy to 
phase-transitions in Statistical Mechanics. 
A general expression for $P$ is derived, and its 
sensitivity to the details of the distribution 
of a single $n_m$ is worked out. These general results are 
verified by the analytical and numerical solution of some 
specific examples.
\end{abstract}

\maketitle

\section{Introduction}

Condensation is the phenomenon whereby
a finite fraction of some quantity, e.g. a particle density, 
concentrates into a small region of 
phase-space, as in the paradigmatic example of a vapor
transforming into a liquid when crossing a phase-transition.
Condensation is observed in a number of different models,
related to magnetic properties \cite{condzrp1}, gravity \cite{bialas1,bialas2,bialas3},
mass transport and other issues 
\cite{condzrp2,condzrp3,condzrp4,condzrp5,condzrp6,condzrp7,condzrp7a,condzrp7b,
condzrp7c,condzrp7d,condzrp8,condzrp9,god1,god2,god3,god4}. 
Despite the prominent role played by molecular interactions in most cases,
condensation can also be observed in {\it non-interacting} systems as,
for instance, in a quantum Bose-Einstein condensate \cite{bec} or in 
classical models such as the spherical model of a ferromagnet \cite{condzrp2}. 
In the above mentioned cases condensation occurs because the condensing quantity --
the particle number in the former example or the spin variance in the latter -- 
is conserved. 
This {\it constraint} acts like an {\it effective} interaction
among the constituents bringing about the transition \cite{effint1,effint2}.
Indeed, condensation does not occur in a non-interacting boson gas -- as
in the case of photons -- which does not conserve the number of constituents.

A different manifestation of condensation is observed when probability
distributions of a fluctuating collective variable ${\cal N}$, 
such as the number of particles in a thermodynamic system, 
are considered. In this case, a fluctuation 
${\cal N}=N$ 
well above the typical value 
can be associated to a condensed configuration of the 
system \cite{effint1,effint2,condfluc1,condfluc2,condfluc3,condfluc4,condfluc5,condfluc6,condfluc7,condfluc8,mars}.
This phenomenon, referred to as {\it condensation of fluctuations},
is not restricted to the particle number ${\cal N}$ but 
was observed for quantities as diverse as energy, exchanged heats, 
particles currents etc... 
\cite{effint1,effint2,condfluc1,condfluc2,condfluc3,condfluc4,condfluc5,condfluc6,
condfluc7,condfluc8,mars}. 
It was shown \cite{effint1,effint2} that in some systems
condensation of fluctuations may occur because, from the mathematical 
point of view,
asking for a specific value ${\cal N}=N$
constraints the system similarly to what a
conservation law does. 

In this paper we study the probability distribution $P$ of
the sum ${\cal N}$ of $M$ non-identically distributed
random variables. We discuss how an interpretation can be provided,
along the guidelines of Statistical Mechanics, in terms of a 
phase-transition between a {\it normal} phase with a vanishing
order parameter $\rho_{\ell}$ and a condensed one with $\rho _{\ell}>0$.
A general expression for the probability $P$ is found and the
radically different behavior of this quantity in the normal and in the
condensed phase are discussed and illustrated by comparing the analytical 
and numerical solution of some 
specific models. In particular, in the condensed phase, the notable
phenomenon of the {\it giant response} -- a dramatic change
of $P$ as the statistical properties of even a single random variable
is modified -- is pointed out.

This paper is organized as follows: In Sec. \ref{themodel} we introduce the 
statistical model that will be studied and set the notation. 
In Secs. \ref{thegas} and \ref{theliquid} its behavior is discussed when 
condensation does not occur and when it does, providing also an example 
by means of an analytically tractable case for identically distributed variables 
(Secs. \ref{exgas} and \ref{exliquid}, respectively).
In Sec. \ref{nonid} the case of non identically distributed variables is
addressed and the phenomenon of the {\it giant response} is discussed (Sec. \ref{giant}).
Some examples are considered in Sec. \ref{exnonid}.
Finally, In Sec. \ref{theconclusions} we briefly summarize and conclude the paper.

\section{The statistical model} \label{themodel}

In order to set the stage, let us consider the
independent variables $n_m$ ($m=1,M$) subject to a probability
$p_m(n_m;K)$, where $K$ is a set of parameters. 
This probabilistic setup is suited to describe 
at a simple level a variety of systems ranging from physics to
chemistry, biology and social sciences. For instance, one can consider
$M$ receptors where $n_m$ ligand particles, like those of 
a pollutant, can be adsorbed, or
$n_m$ electrons populating $M$ atomic levels. 
One can also think of $M$
individuals, or agents, collecting $n_m$ 
{\it resources} with a certain probability $p_m$.
In the former examples
the temperature can be one of the control parameters $K$ but, in general,
others can be present. 

We fix the language by speaking 
of $M$ {\it receptors} hosting a total number 
\be
{\cal N}=\sum _{m=1}^M n_m
\ee
of {\it particles}, with an average value 
$\langle {\cal N}\rangle=\sum _{m=1}^M \langle n_m\rangle$,
where
$\langle n_m\rangle=\sum _nn\,p_m(n;K)$.
For ease of notation the dependence on $K$ will
be often dropped, and
$M$ will be considered large.

We are interested in the probability to observe a total number ${\cal N}=N$ of
particles
\begin{eqnarray}
P(N,M)&=&\sum _{n_1,n_2,\dots,n_M}p_1(n_1)p_2(n_2)\cdots p_M(n_M)
\,\delta_{{\cal N},N}= \nonumber \\
&=&\frac{1}{2\pi i}\oint dz \,e^{M[\ln Q(z,M)-\rho \ln z]},
\label{prob}
\end{eqnarray}  
where, for discrete variables, we used the representation
$\delta _{{\cal N},N}=\frac{1}{2\pi i}\oint dz \, z^{-(N-{\cal N}+1)}$, with 
\be
Q(z,M)=
\left [\prod _{m=1}^M\sum_{n_m}p_m(n_m)
z^{n_m}\right ]^{\frac{1}{M}}
\ee
and $\rho =\frac{N-1}{M}\simeq \frac{N}{M}$ is the particles density.

As explained in \cite{condzrp4,effint1,effint2} the probability distribution $P$
of the fluctuations of the particle number corresponds also to the partition
function of a {\it dual} model where the number of particles is conserved.
For example, with some particular choices of the microscopic probabilities $p$
that will be considered below, such dual model corresponds to specific instances
of the so called urn model (or balls in boxes model) or of a zero-range process. 
This duality, which to the best of our knowledge was never discussed
in connection to the above mentioned models, allows us to borrow a number of 
well established results in this 
research areas to illustrate the behavior of $P$ -- whose properties are here
discussed in a rather large generality -- in some exemplifying cases. 

The following relation holds \cite{condzrp3,bialas3}
\be
P(N,M)=\sum _{n=0}^N \pi(n,N,M)
\label{recorP}
\ee
where 
\be
\pi(n,N,M)=P(N-n,M-1)\,p_M(n)
\label{force}
\ee
is the probability that, when the $M$-th {\it receptor} is added to the
previous $M-1$, $n$ particles are stored in it.
eq. (\ref{recorP}) is a recurrence relation allowing one to determine the probability
distribution of $M$ variables once the one for $M-1$ is known.

Let us discuss the basic mechanisms whereby the recurrence (\ref{recorP}) 
works to build up the full probability $P(N,M)$.
Denoting with $\overline N=M\overline \rho$ the value of $N$
where $P$ is maximum, 
the largest $P(N-n,M-1)$ in eq. (\ref{force}) is the 
one with $n=N-\overline N$. Notice that this term
is present in eq. (\ref{recorP}) only if $N\ge \overline N$. 

\section{Gas phase} \label{thegas}

According to the previous discussion, for $N\le \overline N$, 
$\pi$ has a maximum at a value $n=n_{g}$
which is microscopic and does not scale with $M$
(see the inset of fig.\ref{fig_pi} for a specific example
with $n_g=0$, to be discussed below). 
This is so because the quantities $P(N-n,M-1)$ in eq. (\ref{force})
lower with $n$ (i.e. moving away from the maximum) 
and the same is true for $p_M(n)$, for sufficiently large $n>n_g$
(being $p_M(n)$ normalized).
If the $p_m$'s are monotonously decreasing,
it is $n_{g}=0$. 

A similar setting, with $\pi$ peaked at a microscopic
$n_g$, is found also for $N>\overline N$, when
the largest probability $P(\overline N,M-1)$
is contained in the sum on the r.h.s. of eq. (\ref{recorP}),
but its maximum is tamed by the microscopic probabilities, i.e.
$\lim _{M\to \infty}P(\overline N,M-1)p_M(N-\overline N)=0$. 

The situation with $\pi$ peaked in $n=n_{g}$ is 
physically intuitive: It expresses the fact that, when $M$ is large,
the occupancy $\pi$ of the new receptor (the $M$-th)
is microscopic. We will denote this situation, with a uniformly small
occupation, the {\it normal} (or {\it gas}) phase.

\subsection{An example} \label{exgas}

Let us illustrate these behaviors by considering
a specific example with power-law distribution
\be 
p_m(n)=c\, (n+1)^{-K_m},
\label{zurns}
\ee
where $K_m>1$ and $c$ is a normalization. 

\begin{figure}[h]
\vspace{2cm}
\centering
    %\vbox to 8.5 cm {                                                          
\rotatebox{0}{\resizebox{.85\textwidth}{!}{\includegraphics{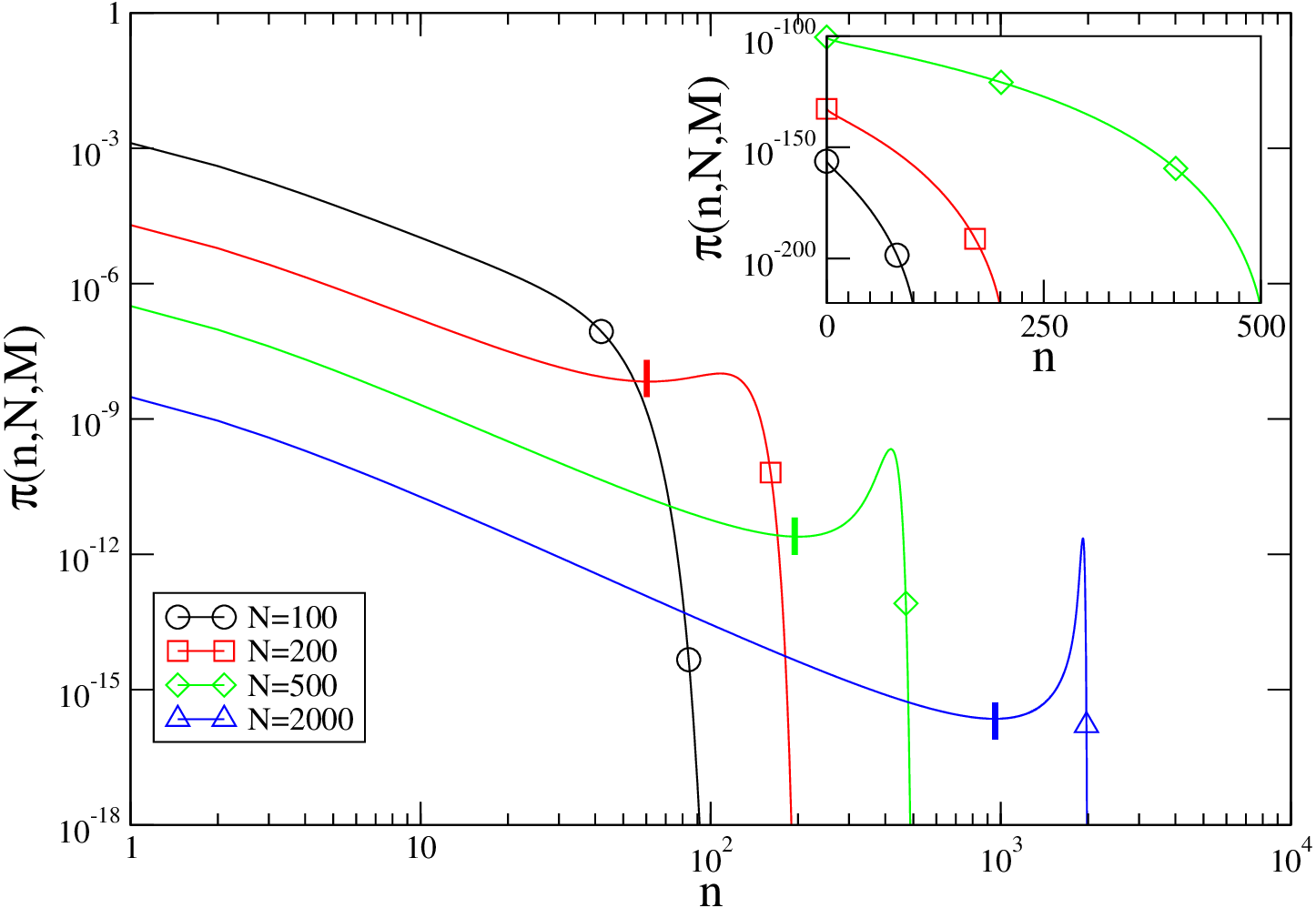}}}

\caption{The quantity $\pi (n,N,M)$ is plotted against
$n$ for the model with probabilities (\ref{zurns}) with a uniform $K_m=K=3>K_c$, 
for $M=246$ and the values of $N$ indicated in the key. The location of $n_{gl}$ 
is indicated with a bold vertical segment. In the inset the same plot is
made for $K_m=K=3/2<K_c$.}
\label{fig_pi}
\end{figure}

We start with the simplest case where $K_m\equiv K$ does not depend on 
$m$ \cite{mars}.
In the inset of fig.\ref{fig_pi}, $\pi$ is plotted
for $K=3/2$ and different choices of $N$. 
Here one observes a sharp peak in $n=n_{g}\equiv 0$, as expected.

In the {\it gas} phase, $P$ can be determined in the large-$M$
limit by a steepest-descent
evaluation \cite{bialas1,bialas2,god1,god2,god3,god4,condzrp3,effint1,effint2,mars,touch} of the integral on the r.h.s. of eq. (\ref{prob}), leading to
\be
P(N,M)\simeq e^{-MR(\rho)},
\label{rate}
\ee
with an $M$-independent rate-function
\be
R(\rho)=-\ln Q[z^*(\rho)]+\rho \ln z^*(\rho),
\label{rrate}
\ee
where $z^*(\rho)$ 
is given by the saddle-point equation
\be
z^*(\rho)\,\frac{Q'[z^*(\rho)]}{Q[z^*(\rho)]}=\rho.
\label{saddle}
\ee

\begin{figure}[h]
\vspace{2cm}
\centering
    %\vbox to 8.5 cm {                                                          
\rotatebox{0}{\resizebox{.85\textwidth}{!}{\includegraphics{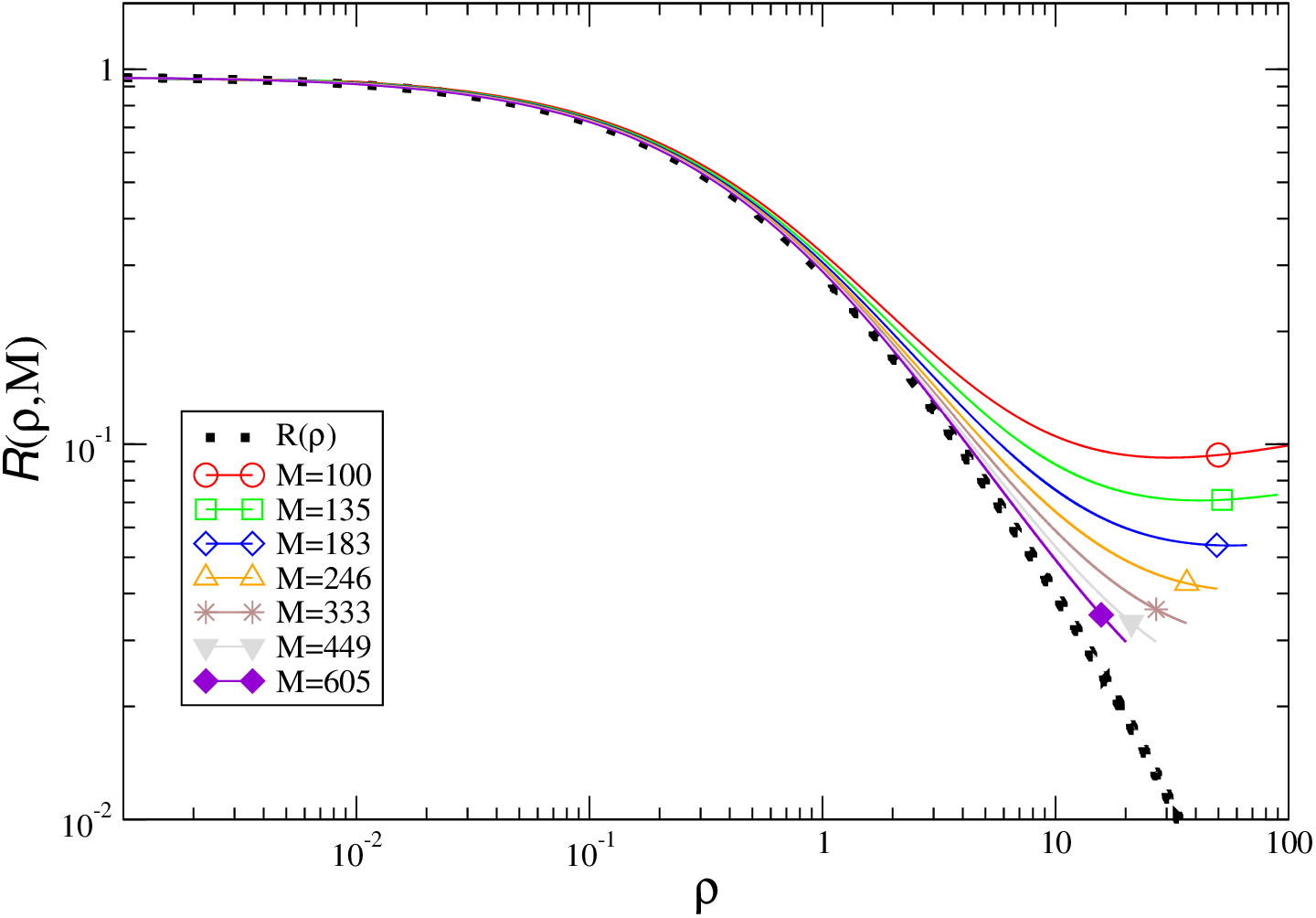}}}
\caption{${\cal R}(\rho,M)$ is plotted against $\rho$ for
for the model with probabilities (\ref{zurns}) with a uniform $K_m=K=3/2<K_c$,
for different values of $M$ in the key. The heavy dashed line is the
asymptotic (for $M\to \infty$) expression $R(\rho)$ 
obtained from  Eqs. (\ref{rrate},\ref{saddle1},\ref{q}).}
\label{fig_rate}
\end{figure}

For the specific example above,
this equation can be cast as
\be
\frac{Li_{K-1}(z^*)}{Li_K(z^*)}=\rho+1,
\label{saddle1}
\ee
where $Li_K(z)$ is the polylogarithm (Jonqui\`ere's function). 
It admits a solution
with $z<1$ for any finite value of $\rho $ if $K\le K_c=2$ \cite{condzrp3,
condzrp7c,bialas1,bialas2,bialas3,god1,god2,god3,god4}. 
$P$ is then expressed 
by Eqs. (\ref{rate},\ref{rrate}) with  \cite{condzrp3,bialas1,bialas2,bialas3,god1,god2,god3,god4}
\be
Q(z^*)=Li_K(z^*).
\label{q}
\ee
The rate-function $R(\rho)$ obtained in this way 
is plotted with a black heavy dashed line in fig.\ref{fig_rate}.
In the same picture the quantity 
\be
{\cal R}(\rho,M)=-\frac{1}{M}\ln P(M\rho,M)
\label{calr}
\ee 
obtained from eq. (\ref{prob}) by exact enumeration, 
is shown for different choices of $M$. One observes that ${\cal R}$ approaches
the asymptotic $M$-independent form $R(\rho)$
as $M$ is increased.
Notice that the convergence is faster at small densities.
It must be recalled that, for $K\le K_c$, the average number $\langle N\rangle$
of particles is not finite, meaning that in the large-$M$ limit fluctuations with large $N$ 
are very likely, as it is reflected by  the vanishing of $R(\rho )$ 
at large densities. However, for finite $M$ such large values of $N$ cannot be
sustained and ${\cal R}$, after reaching a minimum, raises again increasing $\rho$,
thus determining the existence of a most probable value of the fluctuations.
The position of such value is pushed to larger densities by increasing $M$,
providing in this way a gradual convergence, from smaller to larger values of $\rho$, 
of ${\cal R}$ towards $R$.

\section{Condensed phase} \label{theliquid}

A radically different situation occurs
for $N>\overline N$, when $P(\overline N,M-1)$
grows fast enough to give 
$\lim _{M\to \infty}P(\overline N,M-1)p_M(N-\overline N)\ne 0$. 
For the choice (\ref{zurns}) with $K_m=K$, this happens
when $K>K_c$.

In this case the sum in eq. (\ref{recorP}) does not only
take contributions around the microscopic value $n_g$, 
since $\pi $ can be non-negligible up to 
a certain $n=n_{\ell}$ of order $M$.
This is because $\pi$ is triggered by the maximum of $P(N-n,M-1)$
at a value $n=N-\overline N$ which, for given $\rho$, is of order $M$ itself.
Usually $P$ is sharply peaked around the maximum as to give 
$n_\ell \simeq N-\overline N$ for $N$ sufficiently larger than $\overline N$.  
Fig.\ref{fig_pi} shows this 
for the model (\ref{zurns}) with $K=3$: $\pi$
develops a second peak at $n=n_{\ell}\simeq N-\overline N$ 
and only for $n>n_{\ell}$ it drops down to negligible values. 
The properties of this second peak can be described using extreme-value 
statistics \cite{evansmaj}.

The physical interpretation is that, when a value of $N$ larger than the
typical one $\overline N$ is attempted, the occupancy of 
the new receptor $M$ can be either microscopic or macroscopic. 
Then, together with the uniformly
scarcely populated gas phase,
a {\it liquid}, or {\it condensed}  phase coexists characterized by a
single $n_m$ hosting a finite fraction of the $N$ particles. 

From the previous considerations a close resemblance emerges with the problem of
a gas-liquid transition, with $N$ being a control parameter playing the
role of the volume, $\rho_{\ell}=n_{\ell}/M$
an order-parameter and ${\cal H}\sim-\ln \pi $ an
energetic landscape.
Notice that the amount of {\it condensed} vs {\it normal} 
fluctuations, described by $\rho _{\ell}$, depends on $N$: 
The fraction of condensate is absent 
at $N=\overline N$, and increases with $N$.

When condensation occurs eq. (\ref{saddle1}) has no solutions.
Therefore the steepest-descent evaluation of the integral in (\ref{prob})
cannot be carried over straightforwardly as done for the gas phase in 
Sec. \ref{thegas}. In \cite{mars} an upgraded saddle point technique
based on a density functional approach was shown for a case with continuous
variables related to the example (\ref{zurns}).
Another way of proceeding -- still resorting to 
the saddle point technique -- is illustrated in the Appendix.    
Here we prefer to determine the form of $P$ in a different way,
which is discussed now.
Starting for simplicity from equally distributed variables, $p_m(n)\equiv p(n)$,
we assume that particles condensed in a single receptor contribute a fraction
$C(N,M)=(1-a)P$ of the global probability while the others, scattered over the remaining
locations, provide the remaining part $G(N,M)=aP$. The parameter $a$ depends on $N$ in such 
a way that $a=1$ (there is no condensate) at $N\le \overline N$, while $a\to 0$
(all the particles are condensed) for $N\to \infty$.
Casting eq. (\ref{recorP}) as
\be
P(N,M)\simeq \sum _{n=0}^{n_{g\ell}}\pi (n,N,M)+\sum _{n=n_{g\ell}}^{N}\pi (n,N,M)
=G(N,M)+C(N,M),
\label{twoterms}
\ee
where $n_{g\ell}$ is the value of $n$ where $\pi $ is minimum (see fig.\ref{fig_pi}),
allows one to identify $G$ and $C$ as the first and second sum on the r.h.s., respectively.

For $N\gg \overline N$ the peak around $n_{\ell}$ becomes sharper and 
a gaussian approximation for the evaluation of the second term gives
\be
C(N,M)=(1-a)P(N,M)\simeq \sigma P(N_{\ell},M-1)p(N-N_{\ell}),
\label{eqforc}
\ee
where $N_{\ell}=N-n_{\ell}\simeq \overline N$ and 
$\sigma =\sqrt{2\pi [-(\partial ^2\ln \pi/ \partial n^2) _{n=n_{\ell}}]^{-1}}$. 
Assuming that in the large-$N$ limit $\sigma P(N_\ell,M)$ has only a weak dependence on $N$
eq. (\ref{eqforc}) has an approximate solution with
\be
\sigma P(N_{\ell},M-1)\simeq (1-a)\,b_M,
\label{corb}
\ee
and
\be
P(N,M)\simeq b_M\,p(N-\overline N).
\label{secondsol}
\ee
where we have confused $N_\ell$ with $\overline N$ for large $N$ as expressed below eq. (\ref{eqforc}).
Eqs. (\ref{secondsol}) is a general expression for the probability in the
condensed phase for identically distributed variables. A straightforward generalization
to the case of non-identically distributed variables will be discussed below.

The quantity $b_M$ depends on the structure of the microscopic
probabilities. If the $p$'s do not depend on $M$, we can infer $b_M$ by observing that
$\sum _{N=\overline N}^{\infty}C(N,M) $ --
the number of particles in the condensed phase -- is 
of order $M$. Recalling Eqs. (\ref{eqforc},\ref{corb},\ref{secondsol}) one can argue that 
\be
b_M\sim M.
\label{bidim}
\ee

The results (\ref{secondsol},\ref{bidim}) have a very transparent physical 
meaning \cite{condzrp3,tribel}: when condensation occurs a single receptor 
hosts a number $N-\overline N$ of particles with a probability $p(N-\overline N)$.
The factor $M$ is the number of ways to choose such receptor out of $M$,
which is true when the receptors are identical (indeed we will show in Sec. \ref{nonid} 
that eq. (\ref{bidim}) can be violated for non-identically distributed variables).
Notice that our derivation is of a general character and does not rely on any specific
form of the microscopic probabilities $p(n)$.

We emphasize the crucial role played by the $\delta $-function in eq. (\ref{prob}) which,
as mentioned in the introduction, effectively constraints the total particle number
thus invalidating the central limit theorem which would otherwise apply for the problem
at hand, making the condensation phenomenon possible.

Before moving to the
more general case of a non-identical distribution of the $p_m$'s, we 
illustrate all the above with an example.

\subsection{An example} \label{exliquid}

\begin{figure}[h]
\centering
    %\vbox to 8.5 cm {                                                          
\rotatebox{0}{\resizebox{.85\textwidth}{!}{\includegraphics{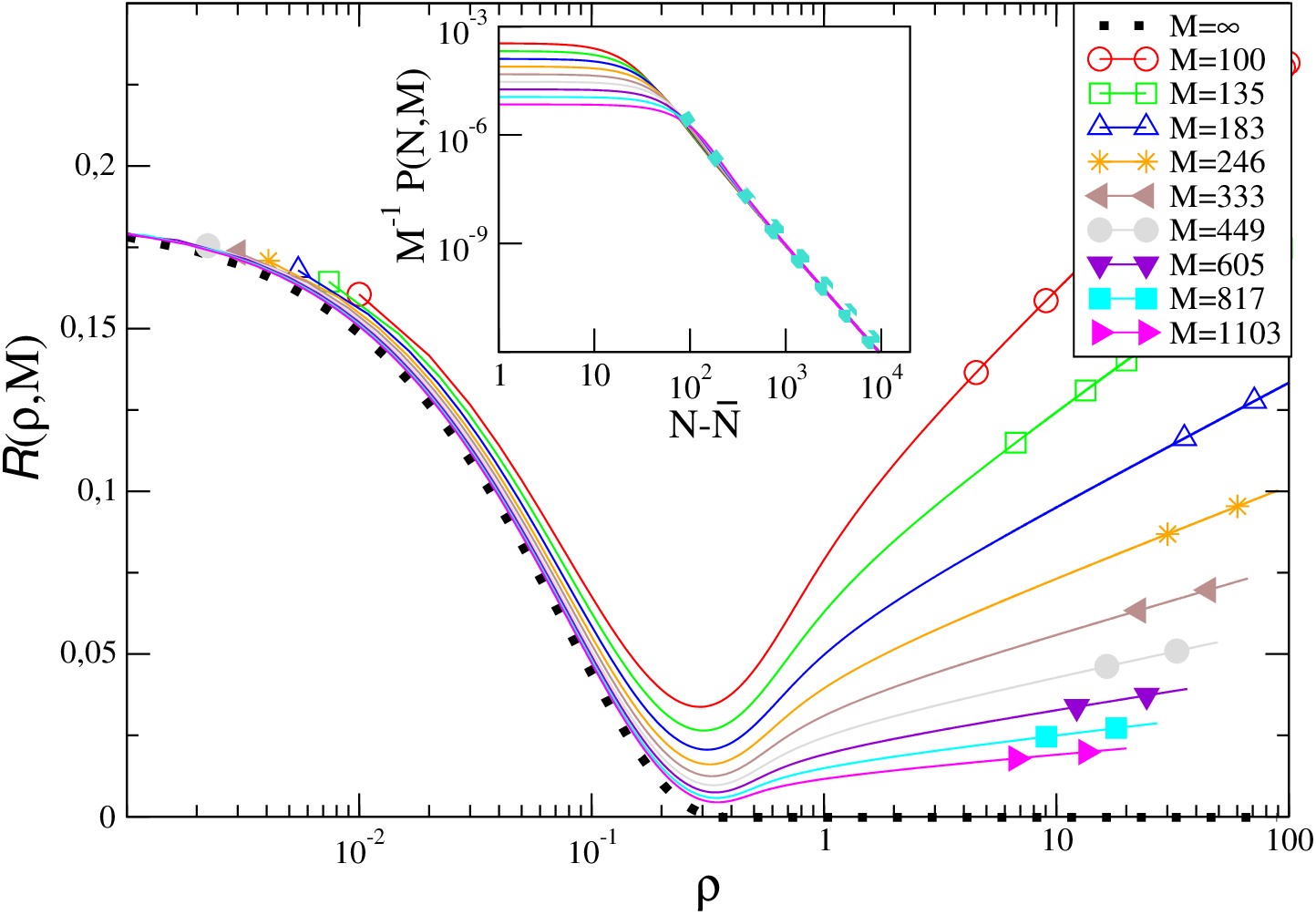}}}

\vspace{2cm}

\rotatebox{0}{\resizebox{.85\textwidth}{!}{\includegraphics{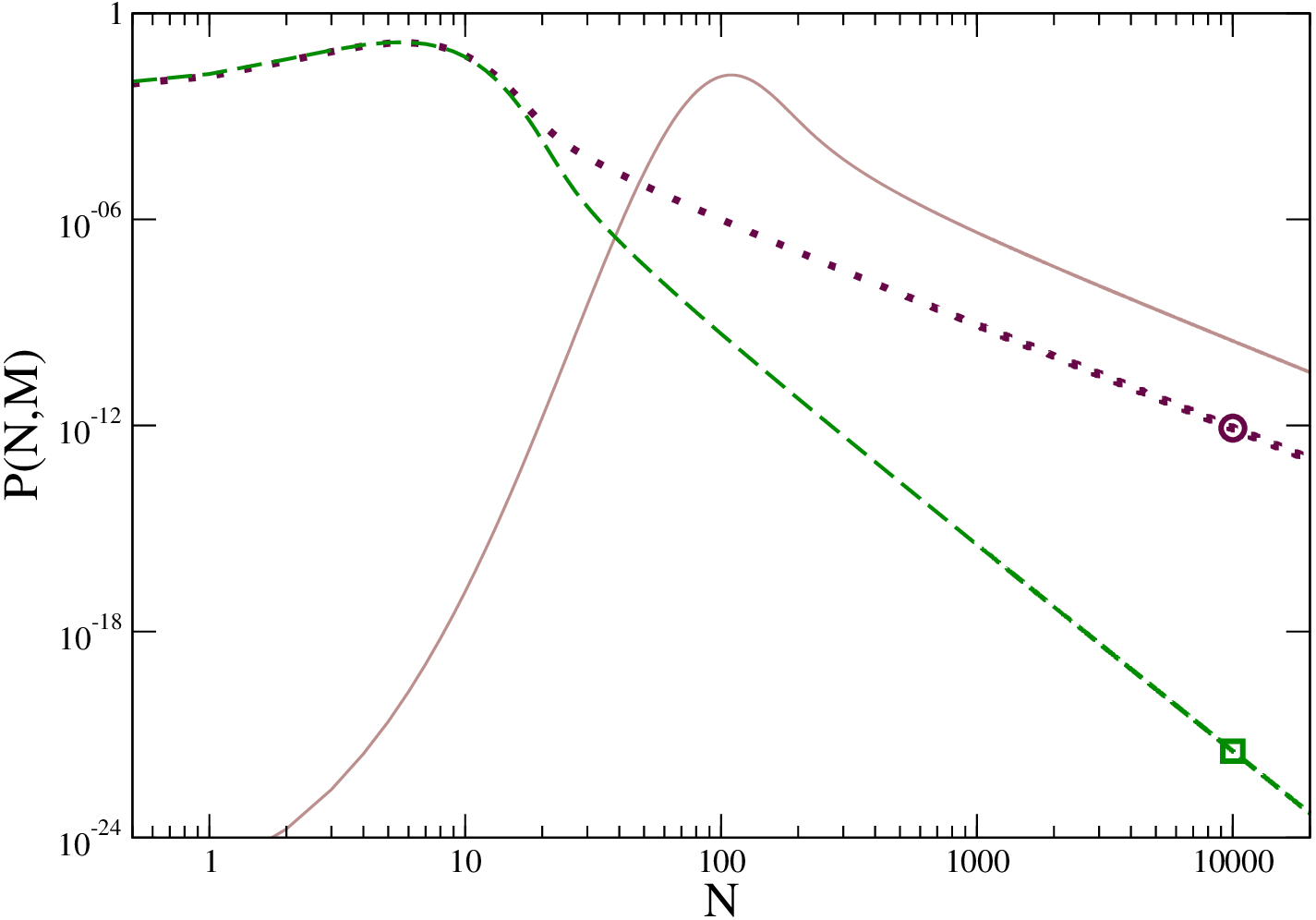}}}
\caption{Upper panel: The function ${\cal R}(\rho,M)$ is plotted against $\rho$
for the model with probabilities (\ref{zurns}) with a uniform $K_m=3>K_c$,
for the values of $M$ in the key. The heavy-dotted black 
line is this quantity for $M\to \infty$, which coincide with $R(\rho)$ 
obtained from  Eqs. (\ref{rrate},\ref{saddle1}) for $\rho \le \overline \rho$
and is identically zero for $\rho \ge \overline \rho$. In the inset 
$M^{-1}P$ is plotted 
against $N-\overline N$. 
The heavy dotted turquoise 
line is the law $(N-\overline N)^{-3}$, i.e. eqs.(\ref{secondsol},\ref{bidim}). 
Lower panel:
$P$ is plotted for $M=333$ and the three different choices (see text) 
i) $K_m\equiv K=3, \,\forall m$, continuous brown, 
ii) $K_m\equiv K=6, \,\forall m$, green dashed with a square and iii)
$K_1=3, K_m=K=6, \,\forall m>1$, dotted maroon with a circle.} 
\label{fig_rate2}
\end{figure}

Let us consider again the distribution (\ref{zurns}) with $K>K_c$. 
It is easy to show that the saddle point solution to 
eq. (\ref{saddle1}) exists only for $\rho \le \overline \rho$ defined by
\be
\overline \rho = \frac{Li_{K-1}(1)}{Li_K(1)}-1
\ee 
($\overline \rho \simeq 0.368433$ for $K=3$).
For
$\rho \le \overline \rho$ only the normal phase exists, 
the steepest descent evaluation of the integral in eq. (\ref{prob}) 
is appropriate, and one arrives at eqs. (\ref{rrate},\ref{saddle1}).
The rate function obtained in this way is shown in the upper panel of fig.\ref{fig_rate2} 
together with the behavior of ${\cal R}(\rho,M)$ [eq. (\ref{calr})] 
which, for $\rho \le \overline \rho$, approaches $R(\rho)$ for large $M$. 

As already discussed, for values of the density larger than $\overline \rho$ 
a straightforward steepest-descent evaluation of the integral in (\ref{prob})
breaks down. In this case
$P$ is not exponentially small in $M$, as
required by eq. (\ref{rrate}), as it can clearly be understood observing in fig. \ref{fig_rate2} that the 
dependence on $M$ does not cancels and ${\cal R}$ keeps
decreasing to zero for any value of $M$.
In this region condensation occurs and,
instead of eqs. (\ref{rrate},\ref{saddle1}),  
the solution (\ref{secondsol},\ref{bidim}) applies.
In order to see this, 
in the inset of fig.\ref{fig_rate2}
we plot $M^{-1}P(N,M)$, since according to eq. (\ref{secondsol},\ref{bidim}) this
quantity ought to be independent of $M$ and proportional to $p(N-\overline N)$. 
As expected, for $\rho \gg \overline \rho$ the form (\ref{secondsol},\ref{bidim}) 
describes the probability with great accuracy for large $N$. 
Notice that in the condensed region the convergence to the asymptotic form 
[eq. (\ref{secondsol},\ref{bidim})] is much faster than the one in the gas phase [to eqs. (\ref{rrate},\ref{saddle1})],
being achieved already for $M\lesssim 100$, a value for which ${\cal R}$ is still quite
different from $R$ in the region $\rho <\overline \rho$.

\section{Non identically distributed variables} \label{nonid}

Now we turn to study the phenomenon of condensation when the
microscopic variables are not identically distributed.
Specific issues of this and related problems have been addressed in 
\cite{condzrp2,condzrp7b,krug,bianconi,igloi1,igloi2,janowsky,derrida,grossinsky}. 
Here we are interest in the derivation of a general form for $P$, generalizing eq. (\ref{secondsol}),
and to discuss the related phenomenon of the {\it giant response} on broad grounds.

When the variables are non-identically distributed one can argue that 
condensation occurs on the most favorable receptor \cite{condzrp2,condzrp7}, 
namely the one with the larger $p_m(n_{\ell})$.
In the example (\ref{zurns}), it is the one
with the smaller $K_m$. Denoting $\overline m$ this term
[i.e. $p_{\overline m}(n_{\ell})> p_m(n_{\ell})$ $\forall m\neq \overline m$],
recalling the physical meaning of $\pi $ in eq. (\ref{force}),
it is clear that a structure like the one in Fig. \ref{fig_rate2}, with a sharp
peak around $n_\ell$, will be present if the recently added receptor is the one where condensation
occurs, namely if $p_M\equiv p_{\overline m}$ in eq. (\ref{force}). 
Then, in order to proceed as in Sec. \ref{theliquid}, we define 
\be
\pi(n,N,M)=P(N-n,M-1)\,p_{\overline m}(n),
\label{newforce}
\ee
which amounts only to the choice of a particular labeling of the receptors.
Proceeding as in Sec. \ref{theliquid} one obtains the following equation
\be
C(N,M)=(1-a)P(N,M)\simeq \sigma P(N_{\ell},M-1)p_{\overline m}(N-N_{\ell}),
\label{eqforc2}
\ee
instead of eq. (\ref{eqforc}), thus arriving at
\be
P(N,M)\simeq b_M\,p_{\overline m}(N-\overline N)
\label{secondsol2}
\ee 
in place of eq. (\ref{secondsol}).
This form of the probability generalize eqs. (\ref{secondsol},\ref{bidim}) to the case
of non-identically distributed variables. Notice that no assumptions on the form of the 
microscopic probabilities $p_m(n)$ has been made also in this case and, therefore,
eq. (\ref{secondsol2}) is expected to hold quite generally. 
Together with eqs. (\ref{secondsol},\ref{bidim}), 
this equation represents the main result of this paper. 
Notice that the dependence on $M$ of $b_M$ can be very different from the one (\ref{bidim})
holding for identically distributed variables. Indeed, when the microscopic probabilities
depend on $M$, the number of particles in the condensed state may not be simply proportional
to $M$, since the $M$-th receptors can promote condensation differently from the previous ones.
An example showing this will be shown in Sec. \ref{exnonid}.
A straightforward consequence of eq. (\ref{secondsol2}) is the phenomenon of the
extreme sensitivity of the global probability $P$ to specific details of the microscopic ones
$p_m$, that we discuss below.

\subsection{Giant response} \label{giant}

It must be stressed that the solutions (\ref{secondsol},\ref{secondsol2}) are 
totally different from the one 
(\ref{rate}), in particular concerning the dependence on $M$.
Indeed, while (\ref{rate}) is exponentially small for large $M$ 
and transforms into a $\delta (N-\overline N)$, 
eq. (\ref{secondsol},\ref{bidim}) shows that in the presence of condensation 
the dependence can be 
as weak as linear in
$M$, signaling the occurrence of anomalously large fluctuations.

Related to that,
an extreme sensitivity of the macroscopic probability 
$P$ to the details of the microscopic ones $p_m$ arises.
In fact, eq. (\ref{secondsol2}) clearly shows that
the distribution of the single variable $\overline m$ 
fully determines the global quantity
$P$. Introducing a {\it susceptibility} $\chi$ --
the shift of the macroscopic probability due to the variation of
a microscopic one -- from eq. (\ref{secondsol2}) one has
\be
\chi (N,m)=\lim _{M\to \infty}\frac{\Delta P(N,M)}{\Delta p_m(N)}=
\left \{ \begin{array}{ll} 
0\,, & \mbox{gas} \\
B_M\, \delta_{m,\overline m}\,, 
&  \begin{array}{ll}\mbox{condens.}\\ (N\gg\overline N),\end{array}
\end{array} \right .
\label{chi}
\ee
where $B_M$ is a constant.
This shows that in the gas phase a shift of one (or even of a finite number) 
of the microscopic probabilities cannot alter the global behavior of $P$, since this is
determined by the synergic contribution of a number $M\to \infty$ of variables,
the situation is profoundly different when condensation occurs.  
In this case $P$ is fully determined by the statistical properties of the most favorable receptor 
$\overline m$. 
Therefore $P$ is independent of the form of all the other $M-1$ variables, whereas a macroscopic effect 
can be determined by altering the statistical properties of the single $p_{\overline m}$. 
As we will show by means of some examples in Sec. \ref{exnonid} this may have
dramatic effects on the form of the probability $P$ in the condensed phase. 

This anomalous susceptibility is reminiscent of the large response induced by gapless
modes \cite{corberi2002}, such as massless Goldstone modes in systems with a spontaneously broken 
continuous symmetry. Actually, starting from equally distributed variables 
the symmetry between the receptors is broken by changing the properties of 
one of them.
However the phenomenon of the giant susceptibility discussed here is more general since
it occurs also when the modification  of the probability of a single variable occurs in a set of
(already) non-identically distributed ones, as will be illustrated by 
the second example of Sec. \ref{exnonid} (fig. \ref{fig_rate3}).
To the best of our knowledge, this remarkable property of the 
{\it susceptibility} (\ref{chi}) was never pointed out before.

\subsection{Examples} \label{exnonid}

The occurrence of condensation in the case of non-identically distributed variables and 
the giant response phenomenon can be illustrated using  
again the probabilities (\ref{zurns}) with
$m$-dependent $K_m$'s. 
The simplest non-trivial choice is when all the $K_m$'s are
equal except one, namely $K_m=K$, $\forall m>1$ while $K_1$ 
can be different from $K$. 

In the lower panel of fig.\ref{fig_rate2} we compare $P$ for 
the three cases i) $K_1=K=3$, ii) $K_1=K=6$ and iii) 
$K_1=3, K=6$.
One sees that the curves relative to the choices ii) and iii)
coincide for $\rho \le \overline \rho$ (i.e. up to the maximum of $P$).
This is because in this region there is no condensation and 
the macroscopic probability is insensitive to a single
$p_m$, eq. (\ref{chi}). 
However, for $\rho >\overline \rho$ the two curves become
totally different and instead case iii) behaves as i),   
apart from a vertical displacement due to the different value of the
constant $b_M$ in eq. (\ref{secondsol2}). 
This shows that a single variable cannot influence the collective behavior
unless it is the one where condensation occurs, in which case a giant response is observed.

The examples considered insofar where based on power-law
probabilities (\ref{zurns}). However, the features above are 
more general and not only restricted to this case.
We show this by considering the exponential form
\be
p_m(n)=c\,\exp\left [-K_m\cdot n^{\kappa _m}\right ],
\label{becchoice}
\ee
where $K_m=\beta \left (\frac{m}{M}\right )^\alpha$ ($\beta $ and $c$ 
are constants), and $\kappa _m$ an $M$-independent exponent. 
This case is interesting also because the microscopic
probabilities $p$ do not depend only on $m$, but also on $M$.
In this case the scaling (\ref{bidim}), which was expected quite generally for 
$M$-independent $p$'s, can -- in principle -- be spoiled, and a general form of
$b_M$ is not available.

\begin{figure}[h]
\vspace{1.5cm}
\centering
    %\vbox to 8.5 cm {                                                          
\rotatebox{0}{\resizebox{.85\textwidth}{!}{\includegraphics{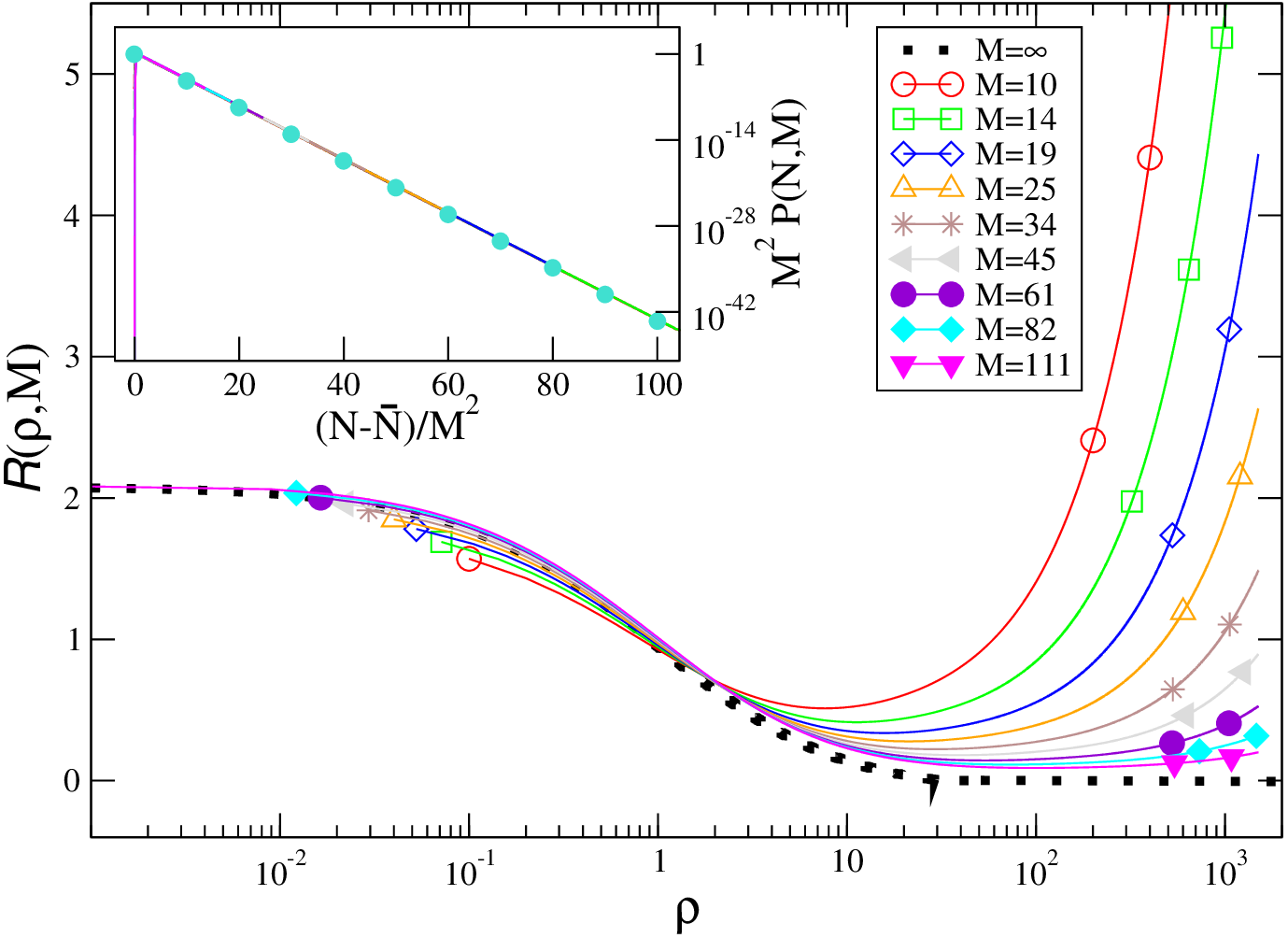}}}

\vspace{2cm}

\rotatebox{0}{\resizebox{.85\textwidth}{!}{\includegraphics{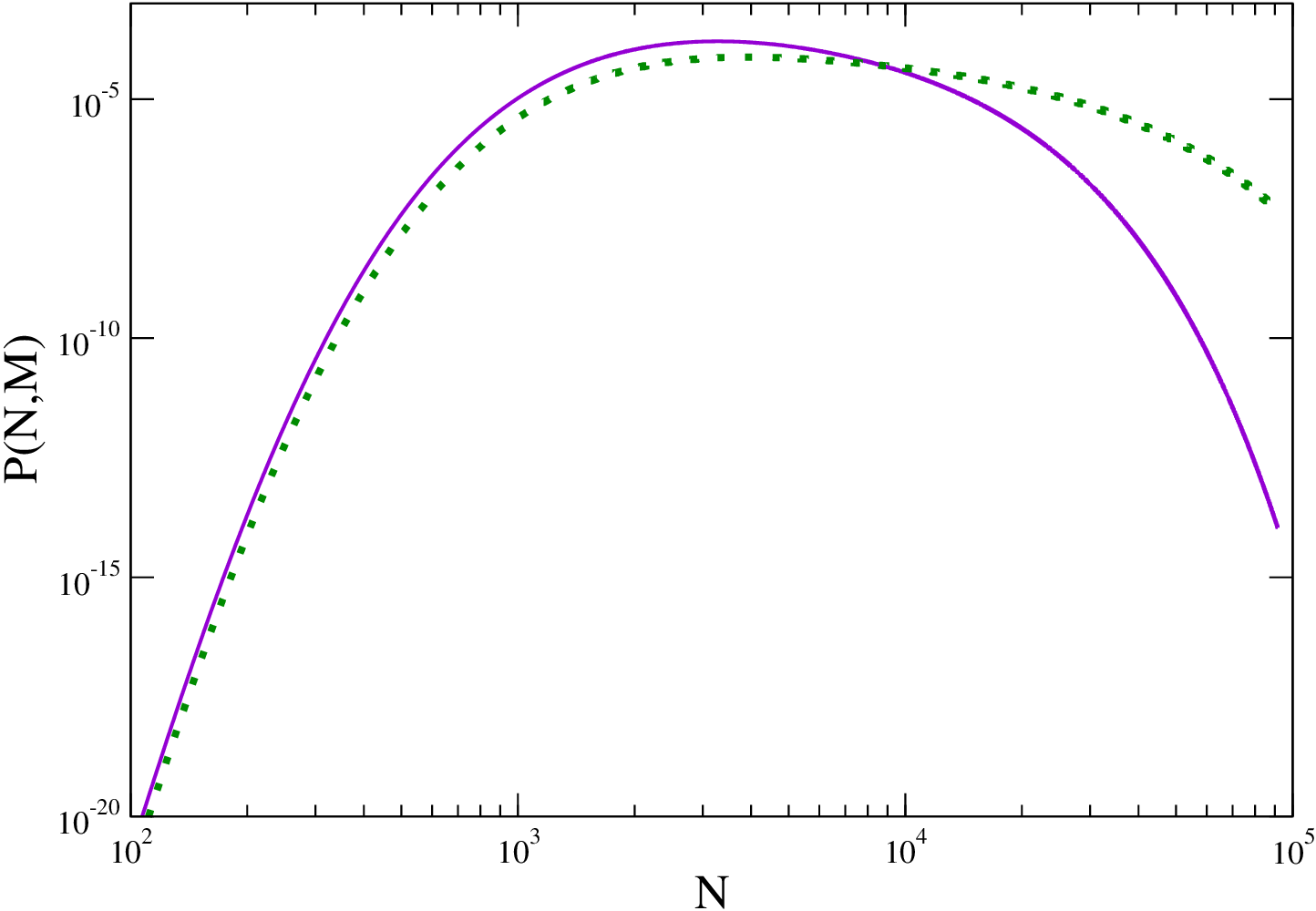}}}
\caption{Upper panel: The function ${\cal R}(\rho,M)$ is plotted against $\rho$ 
for the model with exponential probabilities (see text) with $\beta=1$, 
$\kappa(m)=\kappa\equiv 1$, $\alpha =2$,
for the values of $M$ in the key. The heavy-dotted black 
line is this quantity
for $M\to \infty$, which coincide with $R(\rho)$ 
obtained from  Eqs. (\ref{rrate},\ref{saddle}) for $\rho \le \overline \rho$
and is identically zero for $\rho \ge \overline \rho$. In the inset
$M^{2}P$ is plotted 
against $(N-\overline N)/M^2$. 
The heavy dotted turquoise line is the law $e^{-(N-\overline N)/M^2}$, i.e. eq.(\ref{secondsol2}).
Lower panel: The $P$ of the main figure
with $M=61$ (continuous violet) is compared with the case
(dotted green) with 
$\kappa _1=0.95$, $\kappa _m\equiv\kappa=1$ $\forall m>1$.} 
\label{fig_rate3}
\end{figure}

In order to illustrate the behavior of $P$ with the exponentially distributed microscopic probabilities
we have evaluated it for different
choices of the parameters entering eq. (\ref{becchoice}). 
Starting with a uniform 
exponent $\kappa _m\equiv \kappa =1$, setting $\beta =1$ and $\alpha=2$,
the upper panel of fig.\ref{fig_rate3} shows a pattern of
behavior similar to the case (\ref{zurns}) with
$K>K_c$: for $N\le \overline N$, ${\cal R}$ approaches the form (\ref{rate}) with
a rate function given by Eqs. (\ref{rrate},\ref{saddle}), whereas for $N\gg \overline N$
the determination (\ref{secondsol2}) holds (with $\overline m=1$), implying 
$P(N,M)\sim e^{-\frac {(N-\overline N)}{M^2}}$, as shown 
in the inset. The data collapse is obtained by plotting 
$M^2P(N,M)$ against $(N-\overline N)/M^2$, implying that 
$b_M=bM^{-2}$.
Notice that the approach to the asymptotic form is much faster than for
the fat-tailed probabilities (\ref{zurns}), since already for $M\simeq 100$ one
has a good representation of the large-$M$ form in the range of densities considered,
at variance with what observed in Fig. \ref{fig_rate2}.

The phenomenon of the giant susceptibility is illustrated by comparing 
the case above with the one where we change the distribution of $n_1$  
as to have $\kappa_1=0.95$ and all the remaining ones are left untouched 
($\kappa _m=\kappa,\, \forall \kappa >1$). 
The lower panel of fig.\ref{fig_rate3} shows that,
while in the normal phase $N\le \overline N$
this does not alter $P$ (a residual difference between the two curves is due to the finite value of $M$), 
a dramatic change is produced
in the condensed region $N>\overline N$ because, since $\kappa_1 < \kappa$, the statistical properties
of the condensing variable have been changed.

\section{Conclusions} \label{theconclusions}

In this paper we have discussed the general problem of 
evaluating the probability distribution $P(N,M)$ of 
the sum $N$ of a large number $M$ of micro-variables, not 
necessarily identically distributed.  

We have done this by means of the recurrence relation (\ref{recorP}), which 
provides an analogy with a thermodynamic system where a condensation transition
occurs and the identification of an order parameter $\rho _\ell$. 
Eq. (\ref{recorP}) allows also the derivation of a rather general expression for $P$ 
[eq. (\ref{secondsol2})] which is valid, when condensation occurs, for
finite values of $M$.  
From this expression, computing the susceptibility (\ref{chi}),
the extreme sensitivity of $P$ to the distribution of even a 
single variable was explicitly shown. 

These properties of the probability $P$ have been discussed by means of specific examples
amenable of analytical and numerical computations, including identically and differently
distributed variables, with or without fat-tails and also in the case of a specific dependence 
of the microscopic probabilities $p$ on the number $M$ of micro-variables.

The noteworthy features discussed in this paper are associated to the existence of a 
condensation phenomenon and, therefore, they are
not expected to be only relevant to the large deviations of ${\cal N}$, 
but also to those of different macrovariables, and to apply to a large class of 
problems in Physics and other areas, making the issue considered in this paper a
broad and general research topic.

\acknowledgments
F.Corberi acknowledges financial support by MIUR PRIN 2010HXAW77\_005.

\section{Appendix: Saddle-point evaluation of $P$ in the condensed phase}

In the condensed phase the symmetry between the receptors is broken since, as discussed
in Sec. \ref{theliquid}, a single variable provides a contribution comparable
to all the remaining ones. Let us, without loss of generality, indicate this variable
as being the first, namely $n_1$. In view of that, we re-write the first line
of eq. (\ref{prob}) as follows
\be
P(N,M)=\sum _{n_1} p(n_1) \Omega(n_1,N) 
\label{newprob}
\ee
with 
\be
\Omega(n_1,N,M)=\sum _{n_2,n_3,\dots,n_M}p_2(n_2)\cdots p_M(n_M)
\,\delta_{{\cal N}_1,N-n_1},
\ee
where ${\cal N}_1=\sum _{m=2}^M n_m$.
A solution can be found by making the ansatz that in the condensed phase 
the argument of the sum in eq. (\ref{newprob}) is sharply peaked around
a certain value $n_\ell$, with a certain width $\tilde b$, 
so that it can be evaluated as
\be
P(N,M)=M \, \tilde b \, p(n_\ell)\Omega (n_\ell,N,M)
\label{calpea}
\ee
where 
$\tilde b =\sqrt{2\pi[-(\partial ^2 \ln (p\Omega)/\partial n_1)_{n_1=n_\ell}]^{-1} }$
and the factor $M$ in front of the r.h.s. of eq. (\ref{calpea}) is due to the
$M$ possible ways of choosing the variable denoted by $n_1$ among $M$.
Using eq. (\ref{calpea}) as a starting point, instead of 
eq. (\ref{prob}), one arrives at 
\be
P(N,M)=\tilde b\, M \,p(n_\ell)\, \frac{1}{2\pi i}
\oint dz \,e^{M[\ln Q(z,M)-(\rho -\rho _1)\ln z]},
\label{prob1}
\ee
for large $M$, where now
\be
Q(z,M)=
\left [\prod _{m=2}^M\sum_{n_m}p_m(n_m)
z^{n_m}\right ]^{\frac{1}{M-1}}
\ee
and
$\rho _1=n_\ell/M$ is the condensed particles density.  
The steepest descend evaluation of the integral leads to the saddle point
equation
\be
z^*(\rho)\,\frac{Q'[z^*(\rho)]}{Q[z^*(\rho)]}=\rho-\rho _1.
\label{saddle11}
\ee
In the condensed phase there is always a solution with 
$z=1$ and $\rho _1=\rho - \frac{Q'(1)}{Q(1)}$, and the
evaluation of the integral in eq. (\ref{prob1}) gives 
\be
P(N,M)= b_M  \,p(n_\ell)
\label{prob2}
\ee
with $b_M\sim \tilde b \,M\, e^{M\ln Q(1)}/(2\pi i)=\tilde b \,M/(2\pi i)$ (in the last passage 
we have used $Q(1)\equiv 1$ because of the normalization of the $p$'s). Recalling that 
$n_\ell\simeq N-\overline N$
(see Sec. \ref{thegas}) one recovers the result (\ref{secondsol}) that was obtained
in a different way -- by using the recurrency relation (\ref{recorP}) --
in Sec. \ref{thegas}.

\end{document}